# A Novel Access Control and Privacy-Enhancing Approach for Models in Edge Computing

Peihao Li

Southeast University, Nanjing 211189, Jiangsu, China

**Abstract.** With the widespread adoption of edge computing technologies and the increasing prevalence of deep learning models in these environments, the security risks and privacy threats to models and data have grown more acute. Attackers can exploit various techniques to illegally obtain models or misuse data, leading to serious issues such as intellectual property infringement and privacy breaches. Existing model access control technologies primarily rely on traditional encryption and authentication methods; however, these approaches exhibit significant limitations in terms of flexibility and adaptability in dynamic environments. Although there have been advancements in model watermarking techniques for marking model ownership, they remain limited in their ability to proactively protect intellectual property and prevent unauthorized access. To address these challenges, we propose a novel model access control method tailored for edge computing environments. This method leverages image style as a licensing mechanism, embedding style recognition into the model's operational framework to enable intrinsic access control. Consequently, models deployed on edge platforms are designed to correctly infer only on license data with specific style, rendering them ineffective on any other data. By restricting the input data to the edge model, this approach not only prevents attackers from gaining unauthorized access to the model but also enhances the privacy of data on terminal devices. We conducted extensive experiments on benchmark datasets, including MNIST, CIFAR-10, and FACESCRUB, and the results demonstrate that our method effectively prevents unauthorized access to the model while maintaining accuracy. Additionally, the model shows strong resistance against attacks such as forged licenses and fine-tuning. These results underscore the method's usability, security, and robustness.

## 1 Introduction

With the rapid development and widespread adoption of edge computing, it has demonstrated significant advantages in distributed computing, real-time data processing, and smart device interconnectivity. However, the security and privacy protection of data and models in edge computing environments have become increasingly prominent issues. The distributed nature of deep learning models and data in edge computing scenarios makes them more vulnerable to exploitation by attackers, potentially leading to model theft[1] and data breaches[2]. Relevant study[3] have shown that over 80% of machine learning models deployed in edge applications can be extracted by attackers either directly or through straightforward dynamic analysis techniques, posing severe threats to user privacy and potentially resulting in significant intellectual property losses. These threats not only compromise the security of individual edge devices but may also extend across the entire edge computing network, thereby affecting the overall system's reliability and trustworthiness.

In the context of edge computing, addressing the security of models and data predominantly relies on traditional encryption and authentication methods, such as Public Key Infrastructure (PKI)-based authentication[4, 5] and symmetric encryption for transmission protection[6, 7]. However, these approaches exhibit significant limitations when confronted with the dynamic and heterogeneous nature of edge computing environments. Trabelsi et al.[8] highlight in their research that conventional access control methods struggle with the frequent connectivity changes of edge devices and the adaptability required in heterogeneous network environments. Moreover, model watermarking[9, 10], while effective in marking ownership, is a passive protection technique that fails to prevent unauthorized access and is inadequate for actively safeguarding intellectual property. On the other hand, existing privacy protection techniques face substantial challenges in edge computing scenarios. Traditional methods, such as homomorphic encryption[11, 12] and differential privacy[13, 14], are difficult to deploy in edge environments due to limited computational resources and the high demand for real-time processing. Therefore, there is an urgent need to develop innovative solutions to enhance the security and privacy protection of models and data in edge computing environments.

To address these challenges, we propose a novel intrinsic access control scheme tailored for edge computing, utilizing image style as a licensing mechanism. This approach innovatively integrates image style recognition into the model's internal operational framework, enabling the model to only perform valid inferences on inputs that match specific styles, while invalidating forged styles and arbitrary images. By restricting the model's input data, our method achieves proactive protection of intellectual property, and since the original data must undergo style transfer during use, it also enhances data privacy to a certain extent. Compared to traditional methods, our approach offers greater flexibility and adaptability, making it better suited to address the complex threats in edge computing environments. Extensive experiments conducted on benchmark datasets such as MNIST, CIFAR-10, and FaceScrub demonstrate that our scheme excels in usability, security, and robustness, showcasing its broad application prospects in edge computing scenarios.

The main contributions of this paper are summarized as follows:

- We introduce, for the first time, a novel model access control method that leverages style as a licensing mechanism, and we implement a lightweight style transfer-based license generator suitable for edge computing environments.

- We propose a training loss function and scheme tailored for style license models in edge computing, effectively combining cross-entropy loss, contrastive loss, and style loss to enhance the usability of the style license model.
- We conduct extensive experiments on benchmark datasets such as MNIST, CIFAR-10, and FaceScrub, thoroughly validating the effectiveness of our approach in terms of usability, security, and robustness.

# 2 Proposed Method

## 2.1 Overview

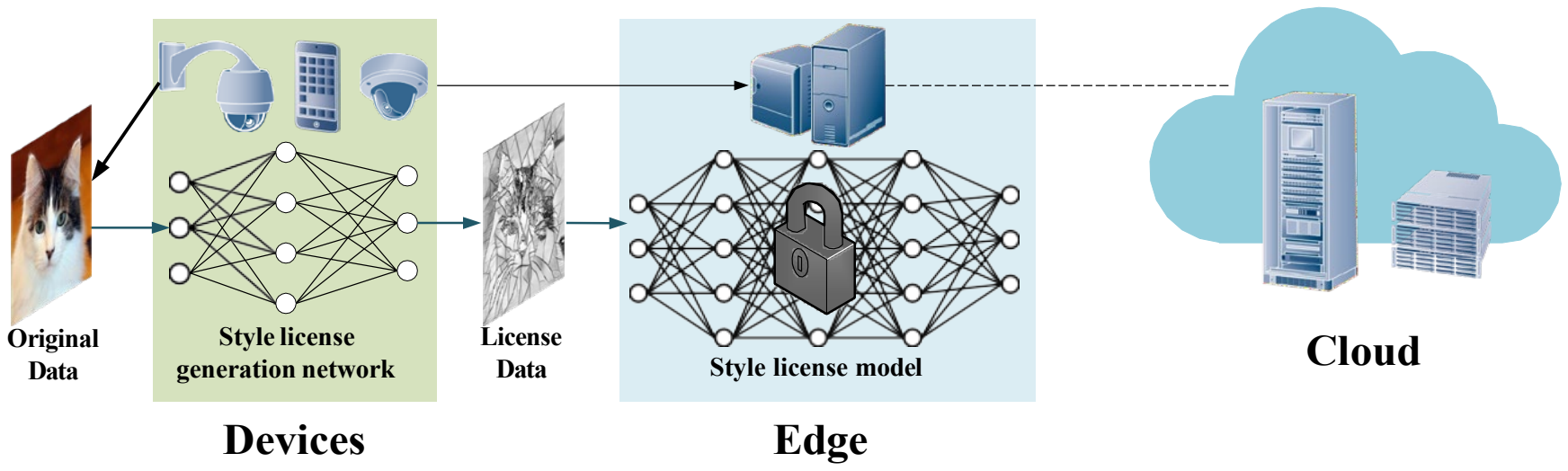

Fig. 1: Inference Process of the Style License Model for Edge Computing

In this section, we provide a detailed description of the novel access control method proposed in this paper. As illustrated in Fig.1, we use an image classification task to exemplify the specific workflow of this approach. First, we train a lightweight style license generation network, which is designed to convert original data collected by terminal devices into stylized license data with a specific style. Next, the license data is transmitted to the edge platform for inference by the model, which has been trained using a combination of style data, original data, and license data based on our proposed style license loss function. This model is characterized by its ability to perform effective inference only on data that matches the authorized style.

Through this method, we achieve two key objectives. First, we embed the access control mechanism directly within the deep learning model deployed on the edge platform. This integration ensures that even if an attacker gains access to the model's parameters and architecture, they would still be unable to use the model without the correct style license, thereby proactively protecting the intellectual property of the model from unauthorized use. Second, in edge computing scenarios, terminal devices are often vulnerable to threats such as man-in-the-middle attacks[15], side-channel attacks[16], and network traffic monitoring[17]. Our approach mitigates these risks by allowing the original data to be converted into the stylized form before transmission, thereby enhancing the privacy of the original data to some extent. The subsequent sections will provide a more detailed explanation of this method.

### 2.2 Lightweight Style Transfer

Given the limited computational resources of edge terminal devices, it is essential to train a lightweight style transfer model to serve as the style license generator for deployment on these devices. After experimentally comparing existing style transfer methods, we selected the Fast Neural Style approach[18] for its lightweight nature and real-time processing capabilities. This method leverages a pre-trained convolutional neural network (CNN) to directly transform input images into a specific style. During training, the generator model is optimized using content and style features extracted from a VGG network. Since the generator model is applied directly in the forward pass of the input image, once training is complete, the model can apply the desired style to any given input image within milliseconds, enabling real-time style transfer.

### 2.3 Style License Loss

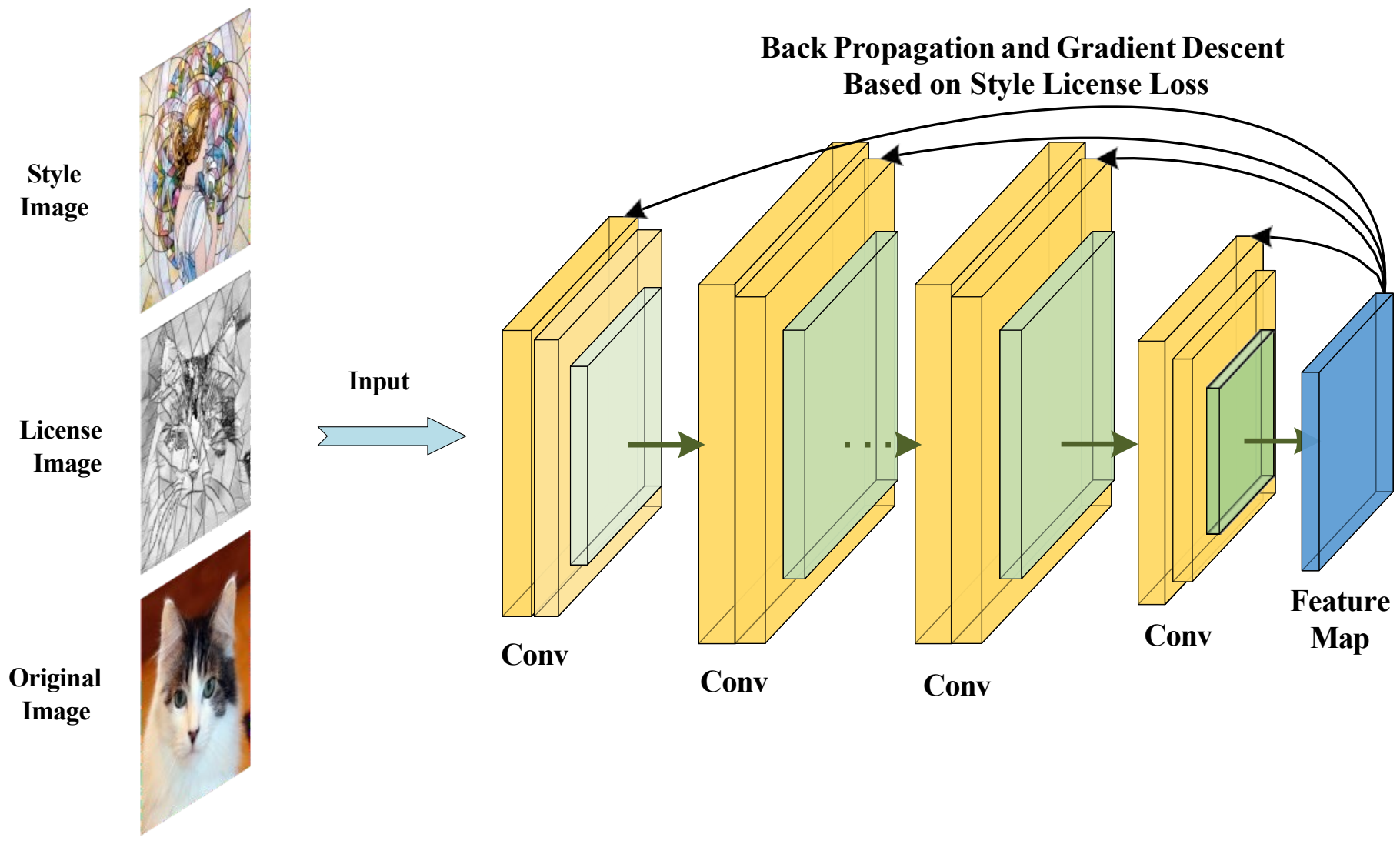

Fig. 2: Demonstration of Style License Model Training Process

To ensure that the model is effective only on style license data while remaining ineffective on any other data, we designed a specialized training scheme and loss function for the style license model. As illustrated in Fig.2, the training process for the style license model involves three datasets:

**Style Dataset:** Images with the licensed style.

**Original Dataset:** Original, non-stylized images.

**Styled-Original Dataset:** Images obtained by converting the original images into the licensed style.

To achieve our desired objective, we developed a custom loss function specifically designed for the convergence of the style license model during training, formulated as follows:

$$L = \alpha L_{\text{Cross-Entropy}} + \beta L_{\text{Contrastive}} + \gamma L_{\text{Style}} \tag{1}$$

The loss function is composed of a weighted combination of cross-entropy loss, contrastive loss, and style loss[19], each serving a distinct purpose in training the style license model. The cross-entropy loss can be represented as follows:

$$L_{\text{Cross-Entropy}} = -y_i \log\left(f\left(x_i^{\text{gen}}\right)\right) \tag{2}$$

where $x_i^{\text{gen}}$ denotes a set of styled-original data, also known as style license data, and $y_i$ represents the corresponding label. This part of the loss function primarily ensures that the model correctly classifies images with the licensed style.

The second part, the contrastive loss, can be expressed as:

$$L_{\text{Contrastive}} = \max\left(margin - d\left(f\left(x_i^{\text{origin}}\right), f\left(x_i^{\text{gen}}\right)\right), 0\right) \tag{3}$$

where $x_i^{\text{origin}}$ represents any set of original data, $x_i^{\text{gen}}$ represents the corresponding styled-original data, and $f(x)$ is the feature output of $x$ in the model. The parameter $margin$ is used to set the minimum distance between the feature outputs of $x_i^{\text{origin}}$ and $x_i^{\text{gen}}$ in the model. The distance metric $d(x, y)$ is a combination of Euclidean distance and cosine similarity, represented as:

$$d(x, y) = \sqrt{\sum_{i=1}^{n} (x_i - y_i)^2} + \varphi\left(1 - \frac{x \cdot y}{\|x\|\|y\|}\right) \tag{4}$$

where $n$ is the dimension of the vectors $x$ and $y$, $\sqrt{\sum_{i=1}^{n} (x_i - y_i)^2}$ represents the Euclidean distance, $x \cdot y$ denotes the dot product of the vectors $x$ and $y$, and $\|x\|$ and $\|y\|$ are their L2 norms, respectively. The parameter $\varphi$ balances the Euclidean distance and cosine similarity within the distance metric. The contrastive loss $L_{\text{Contrastive}}$ is crucial for calculating the feature distance between the original and style license images, ensuring that the model can distinguish between stylized and non-stylized images.

The third part, the style loss, is represented as:

$$L_{Style} = \frac{1}{L} \sum_{l=1}^{L} \left\| G_{\phi}\left(x_j^{style}\right) - G_{\phi}\left(x_i^{\text{gen}}\right) \right\|_F^2 \tag{5}$$

where $L$ denotes the number of layers used to extract style features, $x_i^{style}$ represents a set of style images, and $x_i^{gen}$ represents a set of style license images. $G_\phi(x)$ is the Gram matrix of the feature map at layer $l$ for input $x$. The Gram matrix effectively captures the style information, such as texture and color, from the image data. This part of the loss function ensures that the model effectively learns style features, enhancing its ability to distinguish the licensed style from other inputs.

# 3 Evaluation

In this section, we comprehensively evaluate the proposed style license model. First, we validate the usability of our approach by comparing the accuracy of the style license model with that of traditional models. Then, we assess the privacy protection offered by stylized images by testing them against pretrained traditional image classification models. **All experimental results in this paper are derived based on the MindSpore framework.**

## 3.1 Experimental Setup

Our experiments were conducted using the **MindSpore** framework. We trained baseline models on the MNIST, CIFAR-10, and FaceScrub datasets using VGG16, ResNet18, and MobileNetV2 to evaluate the performance of the style license model. The MNIST dataset contains grayscale images of handwritten digits ranging from 0 to 9, each with a resolution of 28x28 pixels. CIFAR-10 consists of 60,000 color images at a resolution of 32x32 pixels, categorized into 10 classes representing various objects or scenes. Additionally, the FaceScrub dataset comprises 107,818 facial images of 530 celebrities, with all images resized to 3x224x224 pixels for our experiments. VGG16, characterized by its depth and simplicity, contains 16 convolutional layers and 3 fully connected layers, achieving depth through the stacking of 3x3 kernels. ResNet18, which features residual connections, includes 16 convolutional layers and 2 fully connected layers, allowing for the training of deeper networks. MobileNetV2 employs depth-wise separable convolutions combined with linear bottleneck layers to enhance computational efficiency and accuracy, making it well-suited for mobile and embedded systems. In our experiments, we first trained a specific style transfer model using the Fast Neural Style method to serve as the style license generator. This generator was then used to transform the data from the aforementioned benchmark datasets into the corresponding styled datasets, resulting in the creation of the Styled-Original Dataset.

## 3.2 Usability of Style License Model

We evaluated the usability of our approach by assessing the accuracy of the style license model. First, we trained traditional classification models on VGG16, ResNet18, and MobileNetV2 using the MNIST, CIFAR-10, and FaceScrub benchmark datasets, establishing their accuracy as baseline metrics. Subsequently, we trained the style license model using the Styled Dataset, along with the Original Datasets and their corresponding Styled-Original Datasets. We then compared the accuracy of the style license model on both original data and license data against the baseline models. The experimental results are presented in Fig.3.

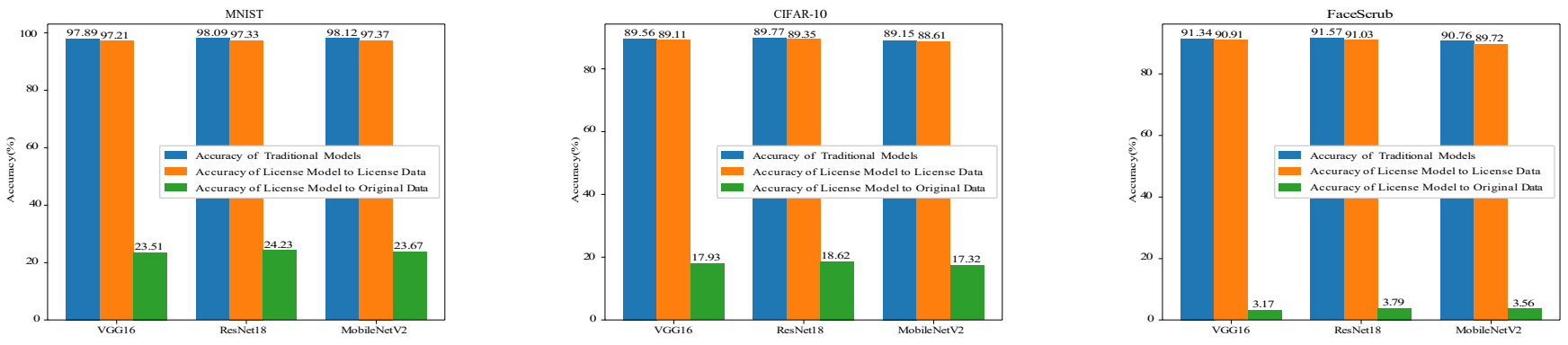

(a) MNIST (b) CIFAR-10 (c) FaceScrub

Fig. 3: The accuracy of style license models on original data and license data compared to that of the baseline models.

As illustrated in Fig.3, the accuracy of the style license models on the license data closely matched the accuracy of the baseline models across all three neural networks and datasets. In contrast, the accuracy of the style license models on the original data was significantly lower than that of the baseline models. This outcome demonstrates that our method effectively maintains model accuracy while preventing unauthorized use, thereby confirming the high usability of the approach.

### 3.3 Privacy Assessment of Stylized Image

To evaluate the effectiveness of our proposed method in protecting data privacy, we compared the accuracy of stylized images versus original images in traditional classification tasks. We trained baseline models using VGG16, ResNet18, and MobileNetV2 on the MNIST, CIFAR-10, and FaceScrub benchmark datasets. The degree of accuracy degradation for stylized images compared to original images in these baseline models was used to assess the extent of privacy protection achieved by the stylization process. The experimental results are shown in Fig.4.

As illustrated in Fig.4, the accuracy of stylized images dropped significantly across all models trained on different datasets compared to the baseline accuracy. Notably, the accuracy for models trained on the CIFAR-10 dataset fell below 40% for most cases. Although models trained on the MNIST dataset maintained over 50% accuracy on stylized data, there was still nearly a 50% decrease compared to the baseline accuracy. This result demonstrates that transmitting stylized images can provide substantial privacy protection for the original data.

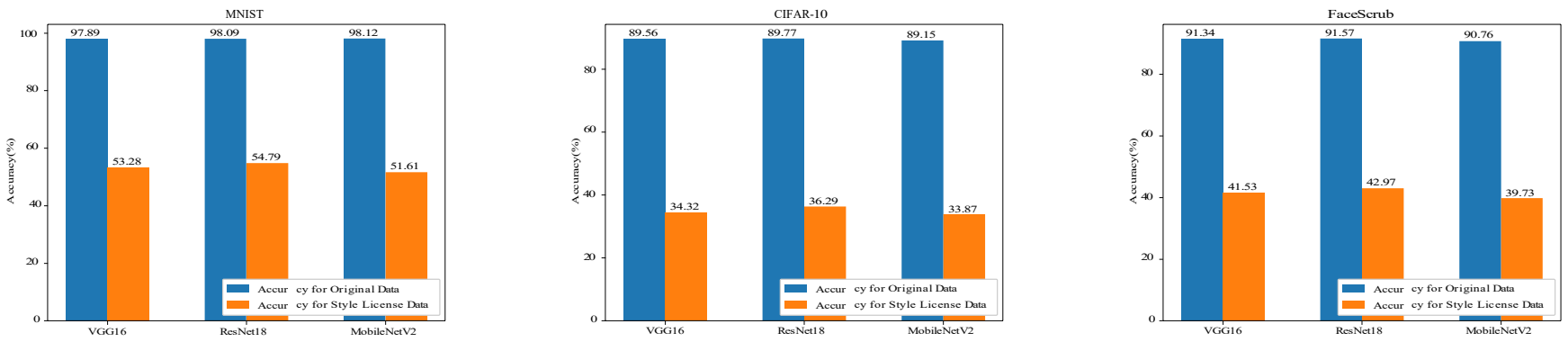

(a) MNIST (b) CIFAR-10 (c) FaceScrub

Fig. 4: Accuracy comparison between original and stylized images in baseline models trained on VGG16, ResNet18, and MobileNetV2 using MNIST, CIFAR-10, and FaceScrub.

## 4 Conclusion

This paper proposes an intrinsic access control method for deep learning models in edge computing environments, based on data style. The approach customizes a license generator for the model and designs a unique training loss function for the style license model, ensuring that the fully trained model can only effectively infer specific styled license data while being ineffective for any other data. By deploying the license generator on the terminal device and the model on the edge platform, the method proactively prevents unauthorized access to the model's intellectual property and enhances data privacy protection on terminal devices. The results demonstrate that the method maintains high accuracy for license data while significantly reducing the model's accuracy on original data, effectively preventing unauthorized access.

## Acknowledgment

Thanks for the support provided by MindSpore Community.

## References

1. T. Orekondy, B. Schiele, and M. Fritz, "Knockoff nets: Stealing functionality of black-box models," in *Proceedings of the IEEE/CVF conference on computer vision and pattern recognition*, 2019, pp. 4954–4963.
2. D. Kolevski and K. Michael, "Edge computing and iot data breaches: Security, privacy, trust, and regulation," *IEEE Technology and Society Magazine*, vol. 43, no. 1, pp. 22–32, 2024.
3. Z. Sun, R. Sun, L. Lu, and A. Mislove, "Mind your weight (s): A large-scale study on insufficient machine learning model protection in mobile apps," in *30th USENIX security symposium (USENIX security 21)*, 2021, pp. 1955–1972.
4. H. Tan, W. Zheng, and P. Vijayakumar, "Secure and efficient authenticated key management scheme for uav-assisted infrastructure-less iovs," *IEEE Transactions on Intelligent Transportation Systems*, vol. 24, no. 6, pp. 6389–6400, 2023.

5. L. Pu, C. Lin, B. Chen, and D. He, "User-friendly public-key authenticated encryption with keyword search for industrial internet of things," *IEEE Internet of Things Journal*, vol. 10, no. 15, pp. 13 544–13 555, 2023.
6. D. Tiwari, B. Mondal, S. K. Singh, and D. Koundal, "Lightweight encryption for privacy protection of data transmission in cyber physical systems," *Cluster Computing*, vol. 26, no. 4, pp. 2351–2365, 2023.
7. C.-I. Fan, C.-H. Shie, Y.-F. Tseng, and H.-C. Huang, "An efficient data protection scheme based on hierarchical id-based encryption for mqtt," *ACM Transactions on Sensor Networks*, vol. 19, no. 3, pp. 1–21, 2023.
8. R. Trabelsi, G. Fersi, and M. Jmaiel, "Access control in internet of things: A survey," *Computers & Security*, p. 103472, 2023.
9. Y. Uchida, Y. Nagai, S. Sakazawa, and S. Satoh, "Embedding watermarks into deep neural networks," in *Proceedings of the 2017 ACM on international conference on multimedia retrieval*, 2017, pp. 269–277.
10. Y. Quan, H. Teng, Y. Chen, and H. Ji, "Watermarking deep neural networks in image processing," *IEEE transactions on neural networks and learning systems*, vol. 32, no. 5, pp. 1852–1865, 2020.
11. R. Gilad-Bachrach, N. Dowlin, K. Laine, K. Lauter, M. Naehrig, and J. Wernsing, "Cryptonets: Applying neural networks to encrypted data with high throughput and accuracy," in *International conference on machine learning*. PMLR, 2016, pp. 201–210.
12. X. Sun, P. Zhang, J. K. Liu, J. Yu, and W. Xie, "Private machine learning classification based on fully homomorphic encryption," *IEEE Transactions on Emerging Topics in Computing*, vol. 8, no. 2, pp. 352–364, 2018.
13. D. Zhang, X. Chen, J. Shi, D. Wang, and S. Zeng, "A differential privacy collaborative deep learning algorithm in pervasive edge computing environment," in *2021 IEEE 20th International Conference on Trust, Security and Privacy in Computing and Communications (TrustCom)*. IEEE, 2021, pp. 347–354.
14. W. Xue, Y. Shen, C. Luo, W. Xu, W. Hu, and A. Seneviratne, "A differential privacy-based classification system for edge computing in iot," *Computer Communications*, vol. 182, pp. 117–128, 2022.
15. M. Conti, N. Dragoni, and V. Lesyk, "A survey of man in the middle attacks," *IEEE communications surveys & tutorials*, vol. 18, no. 3, pp. 2027–2051, 2016.
16. R. Spreitzer, V. Moonsamy, T. Korak, and S. Mangard, "Systematic classification of side-channel attacks: A case study for mobile devices," *IEEE communications surveys & tutorials*, vol. 20, no. 1, pp. 465–488, 2017.
17. M. Abbasi, A. Shahraki, and A. Taherkordi, "Deep learning for network traffic monitoring and analysis (ntma): A survey," *Computer Communications*, vol. 170, pp. 19–41, 2021.
18. J. Johnson, A. Alahi, and L. Fei-Fei, "Perceptual losses for real-time style transfer and super-resolution," in *Computer Vision–ECCV 2016: 14th European Conference, Amsterdam, The Netherlands, October 11-14, 2016, Proceedings, Part II 14*. Springer, 2016, pp. 694–711.
19. X. Huang and S. Belongie, "Arbitrary style transfer in real-time with adaptive instance normalization," in *Proceedings of the IEEE international conference on computer vision*, 2017, pp. 1501–1510.